\documentclass[cup6b,epsf]{cupbook}

\usepackage{epsf}

\def\note #1]{{\bf #1]}}

\def\note #1]{{\bf #1]}}

\begin{document}

\newcommand{\lsun}{L_\odot}
\newcommand{\Msun}{M_\odot}
\newcommand{\gcm}{\rm ~g~cm^{-3}}
\newcommand{\cmc}{\rm ~cm^{-3}}
\newcommand{\kms}{\rm ~km~s^{-1}}
\newcommand{\ergs}{\rm ~erg~s^{-1}}
\newcommand{\ergcc}{\rm ~erg~cm^{-3}}
\newcommand{\ml}{~\Msun ~\rm yr^{-1}}
\newcommand{\mll}{\Msun ~\rm yr^{-1}}
\newcommand{\Mdot}{\dot M}
\newcommand{\la}{\raise0.3ex\hbox{$<$}\kern-0.75em{\lower0.65ex\hbox{$\sim$}}}
\newcommand{\ga}{\raise0.3ex\hbox{$>$}\kern-0.75em{\lower0.65ex\hbox{$\sim$}}}

\author{ROGER A. CHEVALIER\\
Department of Astronomy, University of Virginia,  \\
P.O. Box 3818, Charlottesville, VA, USA}

\chapter{Supernova Remnant and Pulsar Wind Nebula Interactions}

{\it I review several topics in the structure of supernova remnants.
Hydrodynamic instabilities in young remnants may give rise to the
cellular structure that is sometimes observed, although structure in the ejecta
might also play a role.
The presence of ejecta close to the forward shock front 
of a young remnant can be the
result of ejecta clumps or the dynamical effects of cosmic rays.
Slower moving ejecta clumps can affect the outer 
shock structure of older remnants
such as Vela.
Young remnants typically show a circular structure, but often have
a one-sided asymmetry; the likely reasons are an asymmetric circumstellar
medium, or pulsar velocities in the case of pulsar wind nebulae.
In older remnants, asymmetric pulsar wind nebulae can result from
asymmetric reverse shock flows and/or pulsar velocities.
}

\section{Introduction}

Observations of supernova remnants frequently show complex structure that
can have its origin in several ways: structure in the freely expanding
ejecta, structure in the surrounding medium, and the growth of instabilities
that result from the interaction of the supernova with its surroundings.
If we are to infer  properties of the initial explosion from the
supernova remnant, consideration of these various influences is necessary.
Pulsar wind nebulae (PWNe) provide an additional probe inside a supernova
remnant and can lead to an asymmetry because of a pulsar velocity.
Here, I review  studies of these phenomena.

\section{Instabilities in young remnants}

The basic instability that results from the deceleration of the supernova
ejecta by the surrounding medium is related to the Rayleigh-Taylor
instability.
The growth rates for the instability are such that it becomes saturated
early in the evolution of the supernova remnant (Chevalier, Blondin,
\& Emmering 1992).
The instability causes the growth and decay of Rayleigh-Taylor fingers
that only persist a fraction of the region between the forward
and reverse shock waves.
Recent computations of the instability in 3-dimensions show that a
cellular structure develops in which the cells continually grow and
eventually split up (Blondin et al., in preparation).
An estimate of the appearance of the unstable structure by projecting
the emission measure of the gas resembles the cellular structure
observed in Cas A at radio (e.g., Fig. 22 of Rudnick 2002) 
and X-ray (Hughes et al. 2000) wavelengths.
The model assumes that Cas A is interacting with a dense, freely
expanding wind, which appears to be consistent with data on the
shock positions and expansion rates in the remnant (Chevalier \& Oishi 2003).

Tycho's remnant 
is believed to be the remnant of a Type Ia supernova and so is
probably interacting with the constant density interstellar medium.
A comparison of hydrodynamic models
with Tycho's remnant shows that the observed density
structure extends farther toward the forward shock than expected
in the models.
One way to have the structure extend farther toward the forward shock
is to postulate clumps within the supernova ejecta (Wang \& Chevalier 
2001); there is  direct evidence for clumps with specific
chemical compositions in Tycho.
Another way is to have cosmic ray pressure be an important component
of the intershock pressure.
A cosmic ray dominated shock has an adiabatic index approaching 4/3,
yielding a high shock compression; loss of energetic particles from the
shock front can also lead to a larger compression ratio.
Numerical studies of the instability in the case where cosmic ray pressure
is important show a small distance between the forward and reverse shocks
and the instability can grow to the vicinity of the forward shock
(Blondin \& Ellison 2001).

\section{Clumpy/bubbly ejecta}

Clumping of the ejecta can give rise to features that propagate through
the shocked region to create protrusions on the forward shock wave
(Kane, Drake, \& Remington 1999).
A remnant that clearly shows such knots is Cas A, which has fast moving
knots that extend considerably beyond the forward shock front (Fesen 2001).
The knots are observed by their optical emission which implies that
the gas has radiatively cooled.
Radiative cooling at the reverse shock gives rise to denser ejecta, but
does not especially enhance the ability of the instable region to approach
the forward shock front (Chevalier \& Blondin 1995).
The knots must thus be features created in the ejecta.

The origin of the knots is not known.
One possibility is structure created by the expansion of $^{56}$Ni that
is synthesized in the explosion (the Ni bubble effect).
One of the best pieces of evidence for this action is the finding in
SN 1987A that the Fe occupies a large part of the inner volume, but
only a small part of the mass (Li, McCray, \& Sunyaev 1993).
However, knots are observed at high velocities in Cas A (Fesen 2001), in
a region where the Ni bubble effect is not expected to operate.
It may occur at lower velocities, and Blondin, Borkowski, \& Reynolds (2001) 
have modeled the
hydrodynamics of bubble regions in ejecta interacting with a surrounding medium.
The reverse shock front tends to move back rapidly through the
low density bubbles.
The result is added complexity to the structure that is already
affected by instabilities.

The effects of ejecta clumps might also be present in older remnants if
one considers lower velocity clumps ($\la 3000\kms$).
The Vela remnant, with an age of $\sim 10^4$ years, shows evidence for protrusions
at X-ray and radio wavelengths that may be ejecta clumps that are moving
through the outer shock front (Aschenbach, Egger, \& Trumper 1995;
Strom et al. 1995).
The evidence that these may be ejecta clumps include
 the pressure gradient associated with Knot A (Miyata
et al. 2001) and the high pressure present in Knot D (Sankrit, Blair,
\& Raymond 2003).
The clumps that are needed to bring about the protrusions involve a large
density contrast and a relatively large mass,   $\ga 0.01
\Msun$, and their origin
is not clear (Wang \& Chevalier 2002).
The action of the Ni bubble effect is probably not sufficient to give the
required compression.
Another possibility that is applicable to Vela is the sweeping action 
in the supernova interior by 
a pulsar nebula, as can be seen in the Crab Nebula.
The photoionized filaments in the Crab have a density that is $\sim 250$
times the volume averaged density, and neutral gas, if present in such a 
situation, would be even denser.
While the gas that is not affected by the pulsar nebula is decelerated by
the surrounding medium, the pulsar nebula clumps can move out in the remnant.
Although these clumps are dense, the question of whether they have the
properties needed for the Vela protrusion remains.

Another aspect of the protrusions formed by clumps moving through the
outer shock front is that   ring features may be formed in the outer shock.
Wang \& Chevalier (2002) suggested that the X-ray feature RX J0852.0--4622, 
which has been interpreted
as a separate supernova remnant projected on the Vela remnant
(Aschenbach 1998), is actually
a feature in the shell of the Vela remnant.
This view would be negated by clear evidence that a central compact X-ray
source in RX J0852.0 (Pavlov et al. 2001) is associated with the nebula.
If the Vela interpretation is correct, other rings associated with protrusions
might be expected.
In fact, Carlin \& Smith (2002) have found an 
optical emission line ring that appears to
be associated with knot C.
In this case, the shock front is radiatively cooling, so the ring appears
in optical emission as opposed to X-rays.

The marked presence of protrusions in the outer shock 
of a remnant may occur at
a particular phase of a supernova remnant when clumps at the inflection
point in the supernova density profile are able to reach the forward
shock front.
In addition to the Vela remnant, protrusions at the outer shock boundary have
 been found in the remnant N63A in the Large Magellanic Cloud
(Warren, Hughes, \& Slane 2003).

\section{Asymmetric young remnants}

Among the young remnants, a surprising number show evidence for a one-sided
asymmetry in velocity and/or position.
In Cas A, the sphere occupied by the fast knots 
is redshifted by $770\kms$ from the presumed expansion center at 0 velocity
(Reed et al. 1995); the sphere is also displaced from the expansion
center in the plane of the sky (Reed et al. 1995; Thorstensen et al. 2001).
In Kepler's remnant, the velocities of slow knots show mean redshifted
velocities and transverse velocities combining to a speed of
$278\kms$ (Bandiera \& van den Bergh 1991) 
and the X-ray emission is stronger on the N side.
In 0540--69 in the Large Magellanic Cloud, 
the velocities of optically emitting gas around
the pulsar nebula are redshifted by $+370\kms$ from the local LMC
rest velocity (Kirshner et al. 1989), and the pulsar nebula in
G292.0+1.8 is displaced relative to the center of the surrounding
remnant shell in the plane of the sky, corresponding to a transverse
velocity $\sim 770\kms$ if it originated in the center of the shell
(Hughes et al. 2001).

There are probably a variety of reasons for these asymmetries.
In Cas A, Reed et al. (1995) attribute the asymmetry to a higher
density in the surrounding medium on the near side.
This view is supported by the facts that the quasi-stationary
flocculi (shocked cloudlets) are predominantly blueshifted and
the pressure appears to be higher on the near side.
The hydrodynamic model of wind interaction (Chevalier \& Oishi 2003)
gives a specific scenario in which this hypothesis can be 
quantitatively tested.
The model would require a large-scale asymmetry in the presupernova
red supergiant wind.
There are few detailed observations of such winds, but at least in
the case of VY Canis Majoris, the wind shows a complex, asymmetric structure
(Monnier et al. 1999); however, these observations refer to a scale of
0.04 pc, considerably smaller than the 2.5 pc radius of Cas A.

Although an explosion asymmetry is not indicated for Cas A, it is
still interesting to check whether the offset of the recently
discovered compact object in Cas A might be related to the one-sided
asymmetry of the supernova remnant.
The compact object is located to the S of the center of expansion
(Thorstensen et al. 2001), but the center of the remnant ring is to
the W of the center of expansion.
Thus there is no direct evidence that the remnant asymmetry is related to
an impulse given to the compact object.

In the case of Kepler, the asymmetric motions of the slow knots,
which are presumably mass loss from the progenitor star, have
been interpreted in terms of a space velocity of the progenitor
(Bandiera 1987; Bandiera \& van den Bergh 1991).
Attractive features of this scenario are that it can explain the
large distance of Kepler from the Galactic plane (400 pc for a
supernova distance of 1.5 kpc) and the strong X-ray emission to
the N can be explained as interaction with a bow shock structure
(Bandiera 1987; Borkowski, Blondin, \& Sarazin 1992).
However, the proper motion of the X-ray emission shows a faster
rate of expansion than would be expected in this picture (Hughes 1999).
Many features of Kepler's remnant remain to be explained, but
the basic picture of interaction with an asymmetric surrounding
medium appears to apply.

The velocities inferred for the compact pulsar wind nebulae (PWNe)
in 0540--69 and G292.0+1.8 
can be attributed to the velocities given to neutron stars
at birth.
Initially, the PWN expands into the uniformly expanding supernova
ejecta, similar to the Hubble flow.
The inner regions of a supernova may not have a strong density gradient,
so the surroundings of the pulsar are similar to what would be obtained
if the pulsar were not moving.

Although there has been considerable interest in the possibility
that core collapse supernovae have a bipolar structure, there
has been little evidence for such structure in young supernova remnants.
In Cas A, Fesen (2001) found evidence for possible bipolar structure in the
fastest moving knots, along the NE--SW axis aligned with Minkowski's
``jet'' feature.
Sulfur-rich knots preferentially appear along this axis.
From X-ray spectroscopy,
Willingale et al. (2003) also found evidence for bipolar structure,
but along a different axis, the N--S axis.
Overall, the appearance of the bright X-ray and radio emission from
Cas A is quite circular.
One reason for this may be that the interaction region of a supernova
with a circumstellar wind tends to show less asymmetry than the asymmetry
present in the supernova ejecta.
The pressure is higher in the lagging part of the interaction, which
tends to reduce the asymmetry in the interaction region
(Blondin, Lundqvist, \& Chevalier 1996).

A recent study of the stability of standing accretion shocks showed
evidence for a strong $\ell =1$ instability which could influence
core collapse supernovae (Blondin, Mezzacappa, \& DeMarino 2003).
Although a number of young remnants show evidence for an $\ell =1$
asymmetry, there is no clear evidence for an asymmetry in the
ejecta.
More detailed modeling of the sources is needed to disentangle
ejecta asymmetries from other asymmetry mechanisms.

\section{Older Pulsar Wind Nebulae}

The interaction of pulsar nebulae with more evolved supernova remnants,
in which the reverse shock front has moved back toward the center of the
remnant, has been the subject of recent investigations.
The PWN is crushed by the reverse shock front and eventually
re-expands (Reynolds \& Chevalier 1984; van der Swaluw et al. 2001).
This process is strongly unstable if the PWN bubble can be assumed
to behave as a hydrodynamic fluid (Blondin, Chevalier, \& Frierson 2001).
In the actual case, the magnetic field in the PWN may play a role,
although the instability is still expected to operate across magnetic
field lines.
The instability might lead to the development of magnetic filaments,
but further exploration of this topic is needed.

Once the PWN has started to interact with the reverse shocked
material, there are two mechanisms that can lead to asymmetries
in the supernova remnant.
One involves interaction with an asymmetric surrounding medium, so
that the existing PWN is pushed away from its central position in the 
supernova remnant
(Blondin et al. 2001); the other is that the high space velocity
of the pulsar carries it to one side of the remnant.
The radio emitting electrons are typically long-lived compared to
the age of the nebula, so the radio structure shows the positions
of particles injected early in the development of the remnant.
X-ray emitted electrons are typically shorter-lived because of
synchrotron losses, so the X-ray emission is close to the present
position of the pulsar.

There are a couple of ways that these mechanisms can be distinguished.
The clearest is when the proper motion of the pulsar has been
observed so it can be directly checked whether its velocity has
carried it from the center of the remnant.
This is possible for the Vela remnant whose pulsar has an optical
counterpart.
The proper motion vector does not take it back to the
center of the Vela X radio PWN (Bock et al. 1998), so that an
asymmetric interstellar interaction appears more likely for this
case.
Another argument against the pulsar velocity being the sole 
mechanism is if the trail left by the radio emission does not
point back to the center of the remnant.
This is the case for the remnant MSH 15-56 (Plucinsky 1998).
Of course, both pulsar velocity and asymmetric interaction can
be important in one remnant.

In summary, the structure of supernova remnants  involves a
complex set of phenomena which are now being elucidated by
improved observational capability, such as that provided by
the {\it Chandra} X-ray observatory.
Detailed models are needed to sort out the effects of explosion
asymmetries, asymmetries in the surrounding medium, and instabilities.

\bigskip\noindent
{\it Acknowledgements.} I am pleased to present this paper in honor
of the 60th birthday of Craig Wheeler, who has been an inspiration
to the supernova community for several decades.
This work was supported in part by NASA grant NAGW5-13272
and by NSF grant AST-0307366.

\end{document}